\documentclass[a4paper]{article}
\usepackage{epsfig,amsmath,amsfonts,amssymb,setspace,multirow,textcomp, subcaption}
\usepackage[T1]{fontenc}
\usepackage[sort&compress]{natbib}
\makeatletter
\renewcommand{\@oddhead}{\textit{} \hfil}
\renewcommand{\@evenfoot}{\hfil \thepage \hfil}
\renewcommand{\@oddfoot}{\hfil \thepage \hfil}

\makeatother

\renewenvironment{thebibliography}[1]{\begin{oldthebibliography}{#1}\setlength{\parskip}{0ex}\setlength{\itemsep}{0ex}}{\end{oldthebibliography}}
\begin{document}
\fontsize{11}{11} \selectfont
\title{\bf Effects of equatorial chorus wave normal azimuthal distribution on wave propagation}
\author{\textsl{H.\,Breuillard$^1$\thanks{hugo.breuillard@cnrs-orleans.fr}, D.\,I.\,Mendzhul$^{2}$, O.\,V.\,Agapitov$^{1,2}$}}
\date{\vspace*{-10ex}}
\maketitle
%\myHeder{\hspace*{-5ex} \fontsize{8}{8}\selectfont Advances in Astronomy and Space Physics, {\bf 2}, 111+-X (2012)}{}
%\myCopy{\fontsize{8}{8}\selectfont\copyright\hspace*{1ex}\sl H.\,Breuillard, D.\,I.\,Mendzhul, O.\,V.\,Agapitov, 2012}{}
\begin{center}
{\small $^1$LPC2E/CNES and University of Orleans, France \\
$^{2}$STUDIUM, Loire Valley Institute for Advanced Studies,
Orleans-Tours, France\\ $^{3}$Taras Shevchenko National University
of Kyiv, Glushkova ave., 4, 03127, Kyiv, Ukraine\\}
\end{center}
\vspace*{-4ex}
\begin{abstract}
 The non-ducted propagation characteristics of the VLF waves in the inner magnetosphere were studied with respect to their  frequency, source localization, and initial angle $\theta_0$, between the wave-normal and the background magnetic field. The ray tracing software based on multi-components cold plasma approach was developed by use of the IGRF magnetic field model and the GCPM model of plasma density. We described dynamics of the wave-normals direction during its propagation and magnetospheric reflection. We showed that whistler waves can be reflected when lower hybrid resonance frequency becomes greater than the wave frequency: $\omega_{LH}>\omega$. It corresponds to magnetic latitude $\lambda\sim50^\circ$. The simulation results confirmed the inapplicability of the quasi-longitudinal approximation to describe the propagation of magnetospheric whistlers. The simulation results of chorus emissions propagation, which used realistic distributions of waves on the initial parameters was presented. Particularly, we obtained distributions of chorus emission waves in dependence on the wave-normal directions for different magnetic latitudes. It is required for studying diffusive processes in the radiation belts. The results were found to be in a good agreement with the CLUSTER STAFF-SA measurements.\\[1ex]
{\bf Key words:} MHD waves and instabilities, Plasma waves and instabilities, Nonlinear phenomena
\end{abstract}
%\begin{multicols}{2}

\section*{\sc introduction}
\indent \indent The most intense electromagnetic plasma waves
observed in Earth's radiation belts and outer magnetosphere,
discrete ELF/VLF chorus emissions, are characterized by rising and
falling tones in the few hundred to several thousand Hertz frequency
range
\cite[]{burton_origin_1974,tsurutani_postmidnight_1974}.
See also reviews by \cite[]{omura_review_1991} and
\cite[]{sazhin_magnetospheric_1992} and references therein. Critical to
radiation belt dynamics, these emissions have been studied
intensively because they play a crucial role in the acceleration of
energetic electrons in the outer radiation belt. Typically ELF/VLF
chorus emissions are observed near the magnetic equatorial plane in
the dawn and dayside outer magnetosphere \cite[]{omura_review_1991}. They
have attracted special attention recently because they were observed
as waveforms, allowing determination of their wave normal vector
distributions. \cite[]{burton_origin_1974} were the first who
determined chorus normal vector distributions using the search coil
magnetometer aboard OGO5, near the equator, at geosynchronous
altitude. Chorus wave normal directions at $\leq 17^\circ$ magnetic
latitudes and $L$-shells $\sim$7.6 were later studied by
\cite[]{hayakawa_wave_1984}. Analysis of wave normals and Poynting
fluxes for separate emission elements has shown that the emissions
are generated in proximity to the geomagnetic equator and propagate
to higher latitudes in a non-ducted whistler mode
\cite[]{burton_origin_1974,hayakawa_wave_1984} or duct mode
\cite[]{yearby_ducted_2011}. Chorus emissions are usually observed in the
Earth's dawn sector between 23:00 and 13:00\,MLT
\cite[]{tsurutani_postmidnight_1974}. These emissions, which propagate in
the whistler mode, consist of two broad frequency bands on either
side of half local equatorial gyrofrequency $\omega_{ce}$ at the
geomagnetic equator along the magnetic field line on which the waves
are observed \cite[]{tsurutani_two_1977, tsurutani_postmidnight_1974}.
If present, the upper band exists in the $\omega/\omega_{ce} \approx
0.5{-}0.75$ frequency range and contains discrete chorus elements
rising at a few kHz/s. The lower band exists in the $\omega
/\omega_{ce} \approx 0.1{-}0.5$ frequency range and contains both
elements rising at a few kHz/s and diffuse elements. In the inner
magnetosphere, $L$-shell is $\sim$2--6, wave generation onset has
been shown to be associated with substorm electron injections
\cite[]{goldstein_wave_1984}. Chorus in the radiation belts is
believed to be generated through electron cyclotron instability when
the distribution of energetic electrons in the 5 to 150\,keV range
is strongly anisotropic \cite[]{trakhtengerts_generation_1999}. This has been
shown to be the case in recent studies of such waves during electron
injections. Chorus waves in the outer dayside region have received
further attention recently, especially as the THEMIS mission was
able to extend previous observations beyond $L\sim 7$ to $L \sim13$.
Dayside quiet-time chorus cannot be explained in the absence of
injections, however, and remains an area of active study. Other
chorus generation occurs in the local minimum magnetic field regions
near the dayside magnetopause on the magnetic latitude near
$\pm(40^\circ{-}50^\circ)$ \cite[]{tsurutani_two_1977}. The
spatial and temporal dependencies of high-latitude chorus parameters
are considerably different from those of the chorus generated near
the magnetic equator. The frequency range of the high-latitude
chorus is similar to chorus waves generated near the magnetic
equator at $L>10$ but their spectral power distribution exhibits two
maxima: $(0{-}0.15)\omega_{ce}$ and  (0.25--0.30)$\omega_{ce}$.
These emissions are mainly detected within 1--2\,$R_{\rm E}$ from
the magnetopause \cite[]{tsurutani_two_1977}.

Before the Cluster mission, observations of ELF/VLF chorus emissions
were mainly made by single spacecraft, such as ISEE~1 and ISEE~2,
which observed many similar events \cite[]{gurnett_initial_1979}. Recent
Poynting flux and polarization measurements aboard Cluster
spacecraft confirmed that the chorus source is located close to the
equatorial plane \cite[]{agapitov_statistical_2011,agapitov_correction_2012,santolik_central_2005,parrot_source_2003}.
Measurements around the magnetic equator demonstrate the change in
sign of the parallel component of the Poynting vector when the
satellites cross the equator region. Poynting vector flux analysis
indicates that the central position of the chorus source fluctuates
along the background magnetic field within 1000--2000\,km of the
geomagnetic equatorial plane in the timescale of minutes. In studies
of chorus emission generation mechanisms, \cite[]{helliwell_theory_1967} and
\cite[]{trakhtengerts_generation_1999} gave theoretical estimates of the scale
size of the wave generation region. Attempts to estimate the scale
size experimentally were made using coordinated CLUSTER
\cite[]{agapitov_chorus_2010,santolik_central_2005} and
more recently THEMIS \cite[]{agapitov_observations_2011,agapitov_multispacecraft_2011}
observations in the radiation belt region.

Although the magnetospheric reflection of whistler chorus is
discussed and simulated in a number of papers, such as
\cite[]{helliwell_theory_1967} and \cite[]{burton_origin_1974}, experimental
confirmation is scarce. Continuous increase in the angle between the
wave vector and background magnetic field for several cases of
reflected chorus in the outer magnetosphere, based on Ogo~5
measurements, was shown in \cite[]{burton_origin_1974}. In
\cite[]{parrot_source_2003} the Poynting vector and wave normal
directions of chorus waves were analyzed using Cluster STAFF-SA
measurements of spectral matrices with a 4\,s time resolution
\cite[]{agapitov_statistical_2011}. Another Cluster spacecraft observed waves
propagating from the geomagnetic equator region and reflected waves
that reached a lower hybrid resonance reflection at low altitudes
and returned to the equator at another location with a lower
intensity \cite[]{agapitov_chorus_2010}. \cite[]{parrot_source_2003}
corroborated this interpretation using ray tracing analysis.
\cite[]{bortnik_unexpected_2008} demonstrated wave reflection, refraction and
resultant inward radial propagation across $L$-shells and MLT over
several $R_{E}$. The agreement between models and THEMIS data in the
above study \cite[]{bortnik_unexpected_2008} encourages use of the ray
tracing technique to determine evolution and consequences of chorus
waves outside the plasmasphere. To this end it is necessary to
develop a numerical ray tracing code that can reproduce wave
distributions based on direct measurements and then to fill up the
observed distributions by the distributions obtained from
simulations.

In this paper we obtain the distribution of chorus wave parameters,
taking as a basis Cluster observations and to study influence of the
equatorial azimuthal wave normal distribution to wave properties at
high latitudes.

\section*{\sc model description}
\indent\indent In this section, we present the numerical model which
computes ray trajectories of chorus emissions (ELF/VLF whistler
modes) in the inner magnetosphere using the approximation of the
cold collision-less multi-components plasma. The calculation is
carried out making use of realistic plasma density parameters and
pre-selected initial wave distributions.

\begin{figure}
\centering
\begin{minipage}[t]{.99\linewidth}
\centering \epsfig{file = 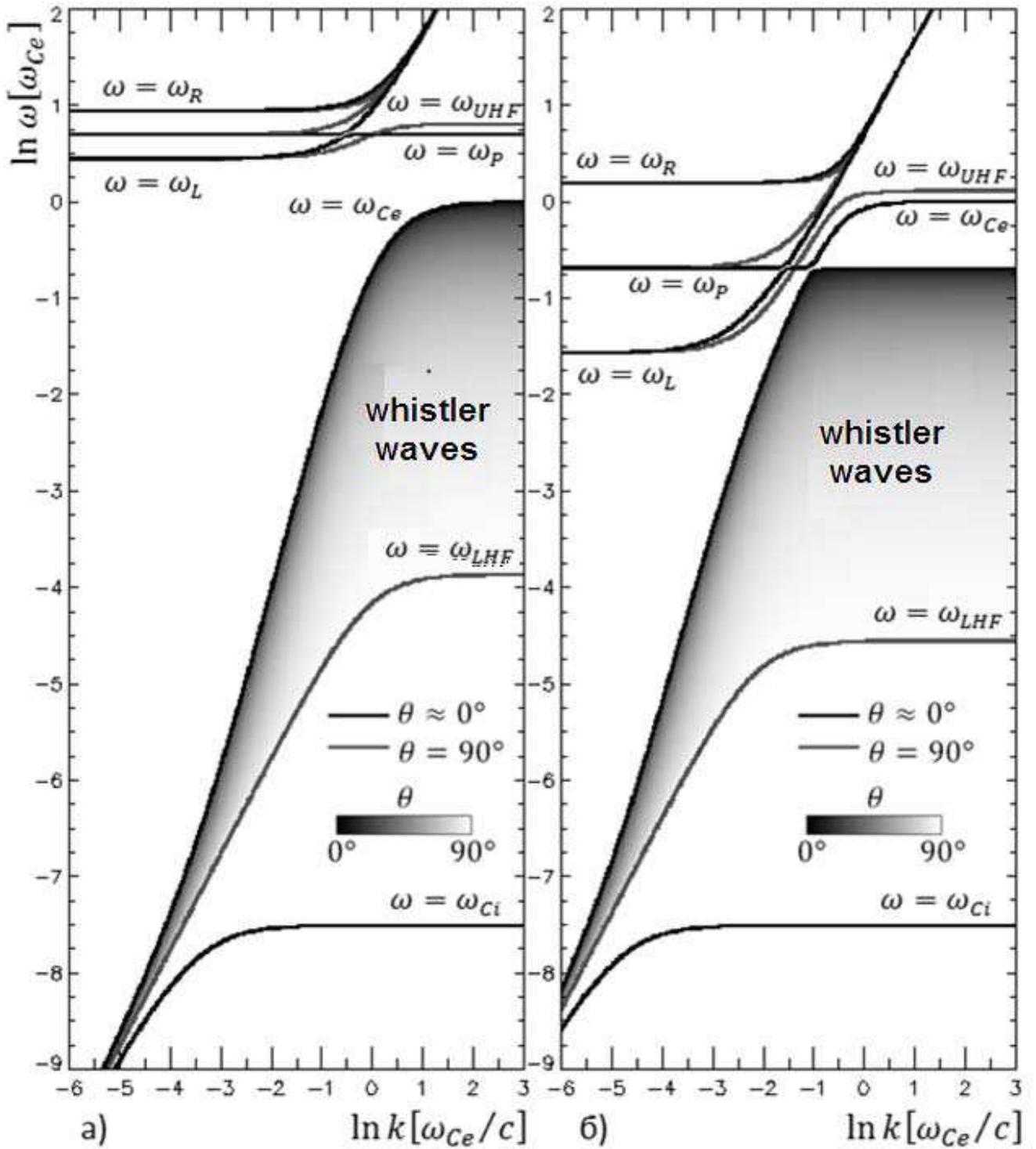,width = 1\linewidth}
\caption{Dispersion of electromagnetic waves for two cases $\omega_{pe}>\omega_{ce}$ (a) and $\omega_{pe}<\omega_{ce}$ (b).}\label{fig1}
\end{minipage}
\end{figure}

The electromagnetic field should be presented in the following form $E,B(\vec{k},\omega,\vec{r},t)e^{-i(\omega t-\vec{k}\vec{r})}$, where field amplitude varies slowly compared with the eikonal $\varphi$ (written as a four-dimensional line integral). In such approximation $\vec{k}=\nabla \varphi(\vec{r})$ and the equations are transformed into a system of eikonal equations. The Maxwell's equations give the equation system for $\vec{E}$ components:
\begin{equation}\label{fr1}
\widehat{M}\vec{E}=\vec{k}(\vec{k}\vec{E})-(\vec{k}\vec{k})\vec{E}+\frac{\omega^{2}}{c^{2}}\widehat{\varepsilon }
\end{equation}
Then, the condition of solvability gives the dispersion relation:
$${M}=\left\|\widehat{M}(\vec{k},\omega ,\vec{r},t)\right\|=0$$
By use of the approximation of cold collisionless plasma in the background magnetic field in the field aligned coordinate system where $\vec{B}_0$ is along $Z$ with the angle $\theta$ between wave normal and the background magnetic field $\vec{B}_0$, the dispersion relation has the form \cite[]{stix_theory_1962}:
\begin{multline}
\left( 1-\frac{\omega ^{2}}{\omega _{pe}^{2}+\omega_{pi}^{2}}\right)\left[\left(\frac{\omega}{\omega_{ce}}\left(1+\frac{\omega_{pe}^{2}+\omega _{pi}^{2}}{c^{2}k^{2}-\omega^{2}}\right)-\right.\right.\\
\left.-\frac{\omega _{ci}}{\omega }\Biggr)^{2}-\left(1-\gamma \right)^{2}\right]+\left[1+\gamma\left(\frac{\omega_{pe}^{2}+\omega _{pi}^{2}}{c^{2}k^{2}-\omega ^{2}}-\right.\right.\\
\left.-1+\gamma \Biggr)-\left( \frac{\omega _{ci}}{\omega }\right) ^{2}\right] \frac{ck\sin
\theta }{c^{2}k^{2}-\omega ^{2}}=M=0,
\end{multline}
where $\omega_{ce}=\dfrac{eB_0}{cm_e}$ is the local electron gyrofrequency, $\omega_{pe}=\dfrac{4\pi{}ne^2}{m_e}$, $\gamma=m_e/m_i=1/1836.15$.
The solutions of the dispersion relation are shown in Fig.\,\ref{fig1} for two cases $\omega_{pe}>\omega_{ce}$ and $\omega_{pe}<\omega_{ce}$ respectively. This dispersion relation is used to solve numerically the equation system for the ray tracing:
\begin{equation}
\frac{d\vec{r}}{dt}=\frac{\partial \omega }{\partial\vec{k}};\quad\frac{d\vec{k}}{dt}=-\frac{\partial \omega }{\partial
\vec{r}},
\end{equation}
in the following form:
\begin{multline}
\frac{d\vec{r}}{dt}=-\left( \frac{\partial M}{\partial \vec{k}}\right) \left( \frac{\partial M}{\partial \omega }\right) ^{-1};\\
\frac{d\vec{k}}{dt}=\left( \frac{\partial M}{\partial\vec{r}}\right) \left( \frac{\partial M}{\partial \omega }\right)^{-1}.
\label{eq0}
\end{multline}
The absolute value of the $\vec{k}$-vector can be obtained from the dispersion relation for whistler waves in a form:
\begin{multline}
\left(\frac{kc}{\omega}\right)^2 = \Biggl(RL\sin^2\theta+PS(1+\cos^2\theta)-\\
-\left[\left(RL\sin^2\theta+PS(1+\cos^2\theta)\right)^2\right.-\\
-\left.4PRL\left(S\sin^2\theta+P\cos^2\theta\right)\right]^{1/2}\Biggr)\times\\
\times\left(2\left[S\sin^2\theta+P\cos^2\theta\right]\right)^{-1},
\label{eq1}
\end{multline}
where $R$, $L$, $P$, and $S$ are the polarization parameters, following the notation by \cite[]{stix_theory_1962}:
\begin{multline*}
R = 1 -\left( \frac{\omega_{pe}}{\omega_{ce}} \right)^2 \left(\frac{\omega_{ce}}{\omega} \right)\left[\left( \frac{\omega_{ce}}{\omega} \right)-1\right]^{-1},\\
  L = 1 - \left( \frac{\omega_{pe}}{\omega_{ce}} \right)^2 \left( \frac{\omega_{ce}}{\omega} \right)\left[\left( \frac{\omega_{ce}}{\omega} \right)+1\right]^{-1},\\
   S = \frac{R+L}{2}, \quad  D = \frac{R-L}{2},\quad   P = 1 -
\left(\frac{\omega_{pe}}{\omega}\right)^2,
\end{multline*}
where $\omega_{pe}$ is the local plasma frequency.

Magnetospheric chorus emissions are commonly observed between
2.5\,$R_E$ and 10\,$R_E$. Near 2.5\,$R_E$ the Earth's core magnetic
field is assumed to have a tilted dipolar structure with the dipole
magnetic momentum corresponding to year 2005 epoch, and an
analytical model includes the contribution of sources external to
the Earth (namely magnetopause, tail and ring currents) at larger
altitudes \cite[]{olson_quantitative_1974}. It is valid for all tilts of the
Earth's dipole axis during rather quiet magnetosphere ($K_p<4$), and
has been optimized for the near-Earth region (from 2\,$R_E$ to
15\,$R_E$). To obtain densities of the magnetospheric plasma species
throughout the entire volume of the inner magnetosphere we use the
Global Core Plasma Model \cite[]{gallagher_global_2000-1} (more detailed description
of the used models see in \cite[]{breuillard_effects_2012}).

\section*{\sc chorus wave normal properties\\ in the outer radiation belt}
\begin{figure}
\centering
\begin{minipage}[t]{.98\linewidth}
\centering \epsfig{file = 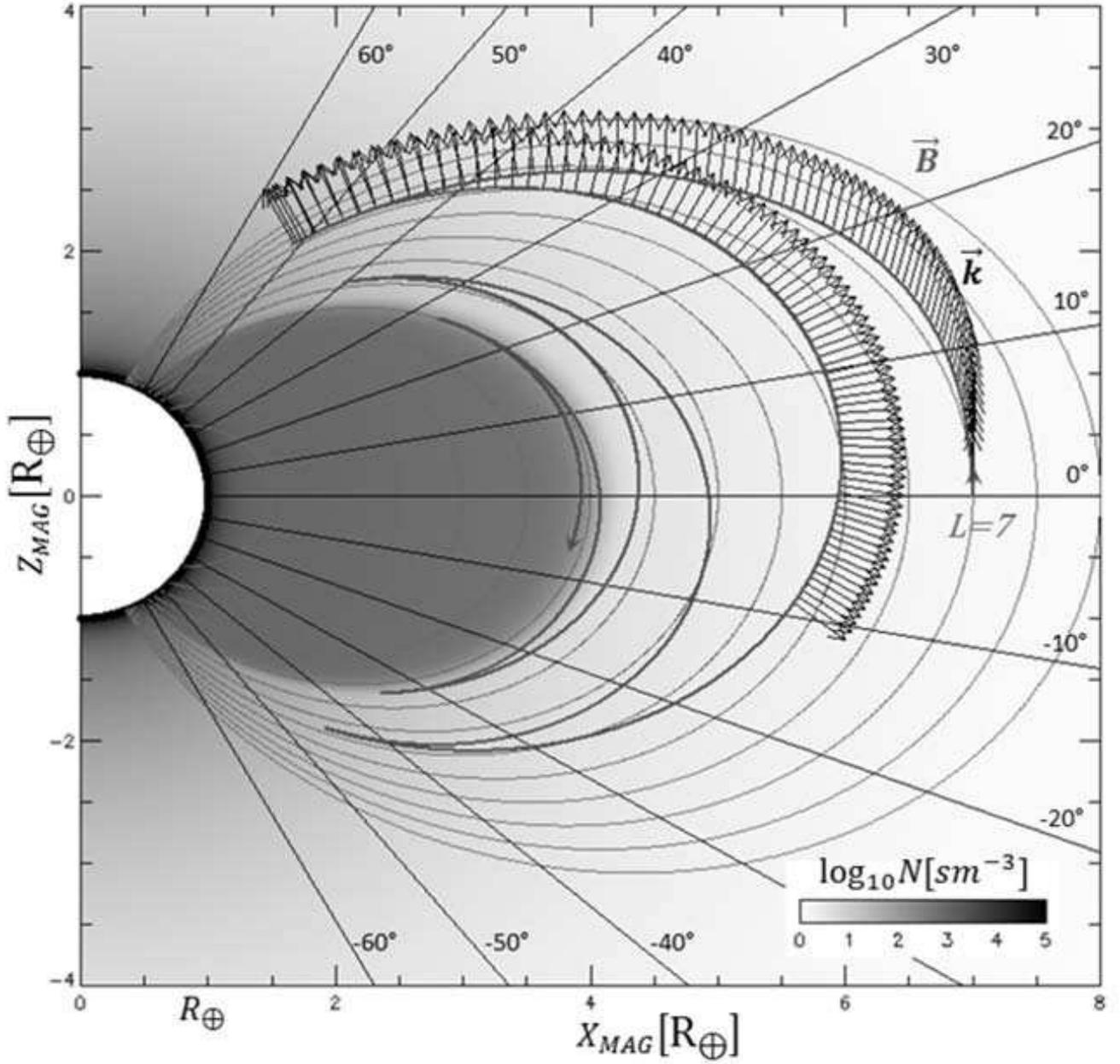,width = .99\linewidth}
\caption{Ray trace of the whistler wave, generated in the vicinity of the geomagnetic equator with wave-normal directed along the background magnetic field. The wave-normal direction along the trajectory is shown by arrows. The model plasma density (the diffusion equilibrium model)  and magnetic field line are shown in grey shades.}\label{fig2}
\end{minipage}
\end{figure}

\begin{figure}[!h]
\centering
\begin{minipage}[t]{.9\linewidth}
\centering \epsfig{file = 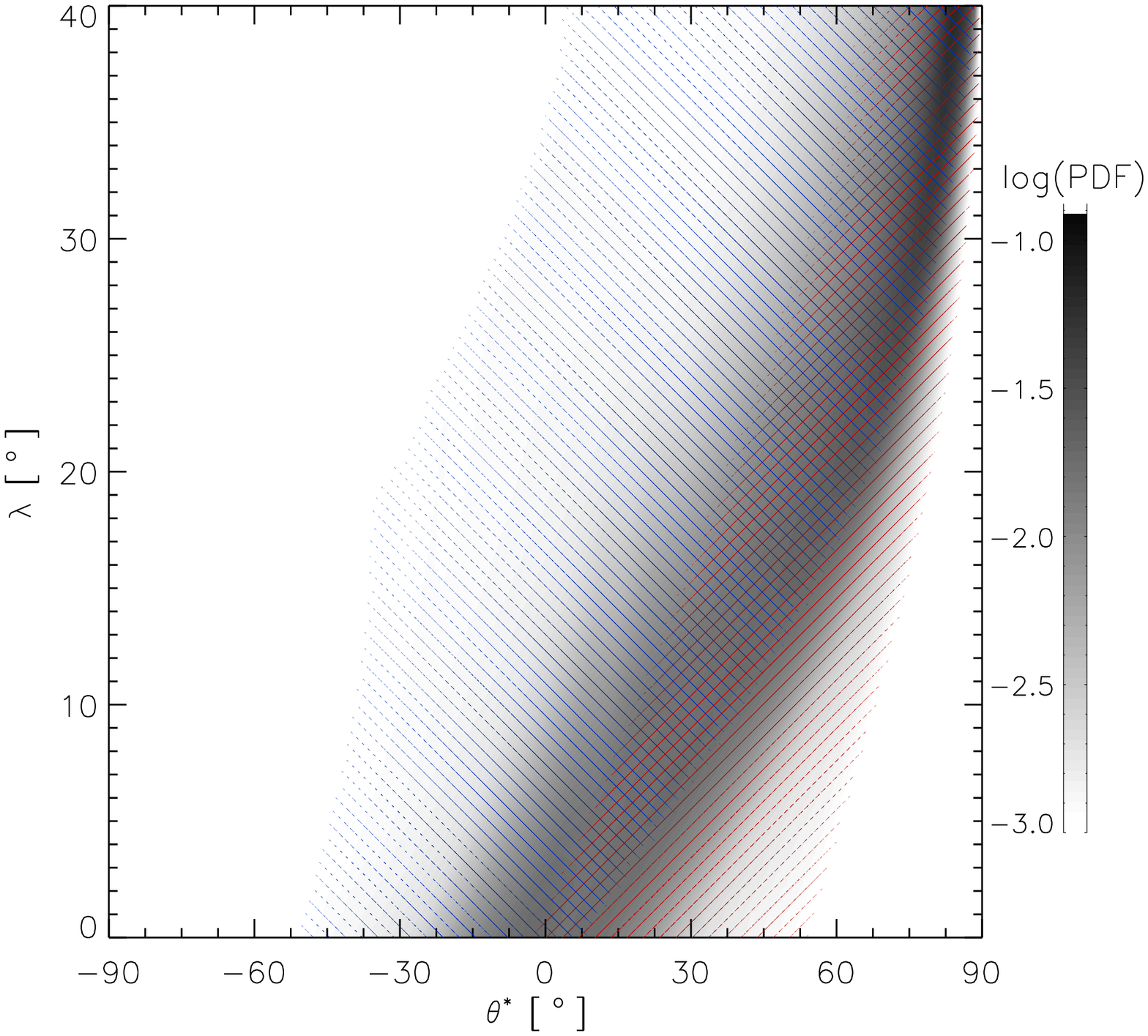,width = 1\linewidth}
\caption{Density probability function of
$\theta^*=\theta\cos\varphi$ as a function of latitude $\lambda$ for
direct chorus waves generated at the equator. The two different ray
populations included in the PDF, i.e. rays for which
$90^\circ<\varphi_0<180^\circ$ and $0^\circ<\varphi_0<90^\circ$, are
indicated by blue and red lines, respectively. The distribution
intensity of each population is represented by the thickness of the
lines.}
\label{fig3}
\end{minipage}
\label{contour_tan}
\end{figure}

\indent\indent The behavior of the numerical distribution of
poleward rays is thus very consistent with the observed distribution
presented in Fig.\,2e from \cite[]{agapitov_statistical_2011}, where the similar
wave normal angle distribution is constructed using the CLUSTER
STAFF-SA measurements from years 2001 to 2009. The good agreement is
confirmed \cite[]{breuillard_effects_2012}. The distributions obtained from
experimental data and as a result of numerical simulations both
exhibit the same tendency, i.\,e. a rapid increase of the mean value
and variance with the growth of $\lambda$. It was shown that
starting with $\vec{k}$ direction close to the direction of the
background magnetic field near the equator, wave-normals diverge
from field aligned direction very fast and already on $\lambda
\approx25^\circ-35^\circ$ became close to the resonance cone
(Fig.\ref{fig2}, \ref{fig3}). When the wave frequency becomes less
than the background lower hybrid frequency
($\omega_{LH}\simeq\left(\frac{1}{\omega_{ce}\omega_{ci}}
+\frac{1}{\omega_{pi}}\right)^{-1}$), whistler wave transforms to
the quasi-electrostatic mode, wave-normal comes to transverse
direction, group velocity changes its sign and wave can be reflected
\cite[]{agapitov_statistical_2011}. From the numerical simulation of chorus
emissions propagation we obtained distributions of chorus emission
waves in dependence on the wave-normal directions for different
magnetic latitudes. The deviation of wave-normal direction from the
direction of the background magnetic field tends to significant
increase of the efficiency of wave-particle interaction in the
radiation belts \cite[]{artemyev_electron_2012, artemyev_distribution_2012}. Due to particle
scattering on chorus waves the electron life-time decreases
\cite[]{mourenas_consequences_2012, mourenas_acceleration_2012}. The wave normal distributions
obtained using hot plasma approximations are described in
\cite[]{breuillard_effects_2012}.

The Fig.\ref{fig3} shows the wave power distribution of $\theta^*$
as a function of $\lambda$, where $\theta^*=\theta\cos\varphi$, for
chorus waves generated at the equator. Thus, the rays that are
initially oriented towards the Earth (i.e.
$90^\circ<\varphi_0<180^\circ$, in blue lines) have a negative
initial $\theta^*$ at the equator, whereas rays initially directed
outward (i.e., $0^\circ<\varphi_0<90^\circ$, in red lines) have a
positive $\theta^*_0$. From this figure it can be seen that rays
quickly defer outward during their propagation, as none is directed
towards the Earth above $\lambda=40^\circ$. Therefore, the majority
of rays generated inward intersect the direction of the local
magnetic field for $\lambda<15^\circ$, which explains the observed
strong population of quasi-parallel waves at these low latitudes.
For large initial angles, these rays
($-30^\circ<\theta^*_0<-15^\circ$) can also propagate quasi-parallel
to the magnetic field up to latitudes $\sim30^\circ$, whereas rays
initially launched outward quickly tend to resonance cone
($\lambda>20^\circ$). However, the two ray populations quickly merge
and this overlapping forms the major part of the distribution,
especially at high latitudes, that propagates obliquely, which is
consistent with Cluster observations \cite[]{agapitov_correction_2012}.

\begin{figure}[!h]
\centering
\begin{minipage}[t]{.9\linewidth}
\centering \epsfig{file = 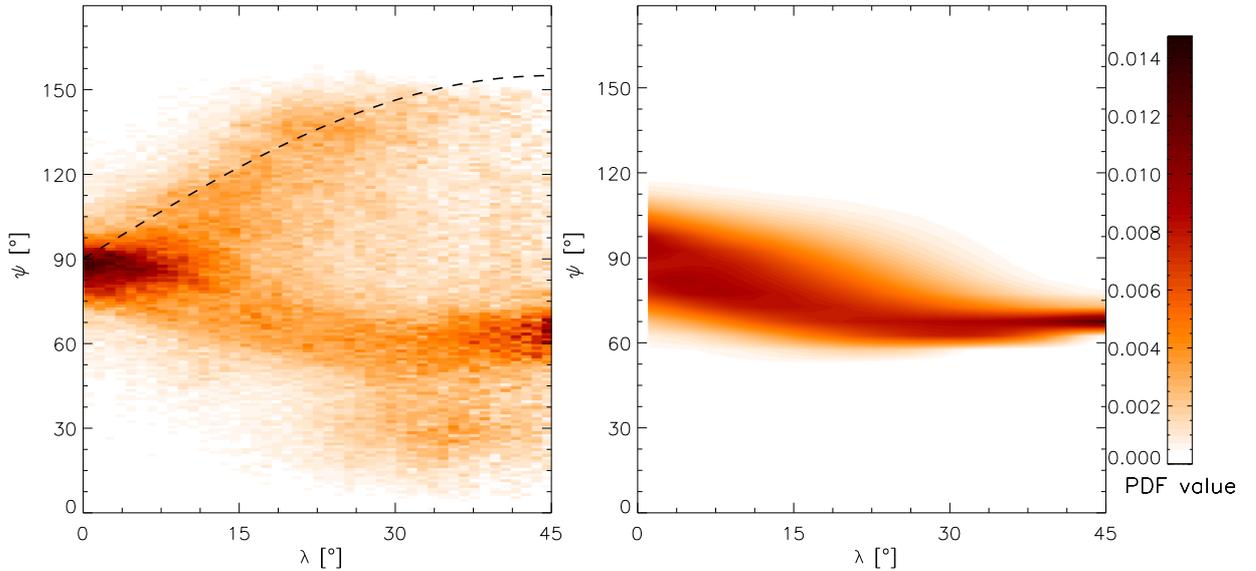,width = 1\linewidth}
\caption{The PDF of $\psi$ distribution ($X$ axis) as a function of magnetic latitude $\lambda$ ($Y$ axis) for all computed rays, using Eq.(\ref{eq0}) as initial weight function. Distributions are obtained from CLUSTER experimental data (left panel) and numerical simulations (right panel).}\label{fig4}
\end{minipage}
\end{figure}

The field aligned population of waves is seen in the spacecraft data
at $\lambda \simeq 20^\circ-30^\circ$. The ray tracing does not
shows such a group of waves. This can be seen also from the angle
$\psi$ between the spacecraft position vector $\vec{r}$ and wave
normal shown in Fig.\,\ref{fig4}. $\psi$ is distributed near
$90^\circ$ at the equator, and from ray tracing decreases with
$\lambda$ to $\sim 60^\circ - 70^\circ$ (Fig.\,\ref{fig4}b). Ray
tracing well reproduces $\psi$ distribution from the spacecraft
observations shown in Fig.\ref{fig4}a. Chorus wave normals   tend
to rotate outward from the Earth due to magnetic field lines
curvature and magnetic field absolute value gradient. The field
aligned direction is indicated with black dashed line and the field
aligned wave population is observed in $\psi$ distribution at
$\lambda \approx 10^\circ-30^\circ$. These field aligned waves
cannot be  obtained from the geometric optics approximation used in
our paper. The Landau damping is more effective for the oblique
waves and wave amplification is more effective for field aligned
waves. The wave amplification and wave damping during the
propagation can result in the observed differences between
spacecraft observations and numerical simulations and they have to
be taken into account, that will be the subject of the future work.

\vspace*{1ex}

\section*{\sc conclusions}
\indent\indent In this paper we present a ray tracing technique for
the cold magnetized multi-components collision-less plasma that has
been developed including realistic plasma density and magnetic field
models. Ray tracing was carried out assuming that the wave source is
situated at equator. The initial distribution of waves in
frequencies and in $\vec{k}$-vectors in the source region was chosen
to correspond to statistical distribution dependencies on
$\vec{k}$-vectors and frequencies obtained from observations. To
this end the weight functions corresponding to distributions
inferred from observations were applied to initial set of rays
\cite[]{breuillard_effects_2012}.

Then, making use of our numerical database we show that it is
possible to reconstruct chorus wave-normal distributions as a
function of magnetic latitude using weight functions on equator
(obtained from experiment). The results of our calculations are in
good agreement with statistical distributions found using ten years
observational data measured onboard Cluster spacecrafts in
\cite[]{agapitov_correction_2012, agapitov_statistical_2011}.

Chorus wave normal angle distribution at the equator peaks at
$\theta \approx 15^{\circ}-17^{\circ}$ and $\theta$ tends to
increase fast with latitude reaching the resonance cone at
$\lambda\approx30^{\circ}-40^{\circ}$ and can be reflected at
$\lambda\approx40^{\circ}-45^{\circ}$ where the wave frequency
becomes less than the local lower hybrid frequency. Chorus wave
normals tend to rotate outward from the direction to the Earth due
to the background magnetic field lines curvature and gradients of
magnetic field and plasma density. The differences of distributions
obtained from the spacecraft observations and from ray tracing can
be explained by the Landau damping of the wave during the
propagation which is much more effective for oblique wave normal
angles. This damping tends to decrease amplitude of oblique chorus
waves and results in the  concentration of the wave normal
distribution near field aligned direction at
$\lambda\approx20^{\circ}-35^{\circ}$. Therefore the existence of
oblique wave population in the outer radiation belts tends to more
effective electron scattering \cite[]{artemyev_electron_2012} and decreasing of
electron life-time, especially in the $L$-shell range from 3 to 5.5
\cite[]{mourenas_timescales_2012}.

\bibliographystyle{natbib}

%\bibliography{Full}

\begin{thebibliography}{28}
\expandafter\ifx\csname natexlab\endcsname\relax\def\natexlab#1{#1}\fi
\expandafter\ifx\csname url\endcsname\relax
  \def\url#1{\texttt{#1}}\fi
\expandafter\ifx\csname urlprefix\endcsname\relax\def\urlprefix{URL }\fi

\bibitem[{Agapitov \emph{et~al.}(2010)Agapitov, Krasnoselskikh, Zaliznyak,
  Angelopoulos, Le~Contel, and Rolland}]{agapitov_chorus_2010}
Agapitov, O., Krasnoselskikh, V., Zaliznyak, Y., Angelopoulos, V., Le~Contel,
  O., and Rolland, G. (2010).
\newblock Chorus source region localization in the {Earth}'s outer
  magnetosphere using {THEMIS} measurements.
\newblock In \emph{Annales {Geophysicae}}, volume~28, pages 1377--1386.
\newblock \urlprefix\url{https://hal-insu.archives-ouvertes.fr/insu-01180462/}.

\bibitem[{Agapitov \emph{et~al.}(2011{\natexlab{a}})Agapitov, Krasnoselskikh,
  Dudok~de Wit, Khotyaintsev, Pickett, SantolÃ­k, and
  Rolland}]{agapitov_multispacecraft_2011}
Agapitov, O., Krasnoselskikh, V., Dudok~de Wit, T., Khotyaintsev, Y., Pickett,
  J.~S., SantolÃ­k, O., and Rolland, G. (2011{\natexlab{a}}).
\newblock Multispacecraft observations of chorus emissions as a tool for the
  plasma density fluctuations' remote sensing.
\newblock \emph{Journal of Geophysical Research: Space Physics (1978â€“2012)},
  \textbf{116}(A9).

\bibitem[{Agapitov \emph{et~al.}(2011{\natexlab{b}})Agapitov, Krasnoselskikh,
  Khotyaintsev, and Rolland}]{agapitov_statistical_2011}
Agapitov, O., Krasnoselskikh, V., Khotyaintsev, Y.~V., and Rolland, G.
  (2011{\natexlab{b}}).
\newblock A statistical study of the propagation characteristics of whistler
  waves observed by {Cluster}.
\newblock \emph{Geophysical Research Letters}, \textbf{38}(20).

\bibitem[{Agapitov \emph{et~al.}(2011{\natexlab{c}})Agapitov, Krasnoselskikh,
  Zaliznyak, Angelopoulos, Le~Contel, and Rolland}]{agapitov_observations_2011}
Agapitov, O., Krasnoselskikh, V., Zaliznyak, Y., Angelopoulos, V., Le~Contel,
  O., and Rolland, G. (2011{\natexlab{c}}).
\newblock Observations and modeling of forward and reflected chorus waves
  captured by {THEMIS}.
\newblock \emph{Annales Geophysicae-Atmospheres Hydrospheresand Space
  Sciences}, \textbf{29}(3), 541.

\bibitem[{Agapitov \emph{et~al.}(2012)Agapitov, Krasnoselskikh, Khotyaintsev,
  and Rolland}]{agapitov_correction_2012}
Agapitov, O., Krasnoselskikh, V., Khotyaintsev, Y.~V., and Rolland, G. (2012).
\newblock Correction to ???{A} statistical study of the propagation
  characteristics of whistler waves observed by {Cluster}???
\newblock \emph{Geophysical Research Letters}, \textbf{39}(24).

\bibitem[{Artemyev \emph{et~al.}(2012{\natexlab{a}})Artemyev, Agapitov,
  Breuillard, Krasnoselskikh, and Rolland}]{artemyev_distribution_2012}
Artemyev, A., Agapitov, O., Breuillard, H., Krasnoselskikh, V., and Rolland, G.
  (2012{\natexlab{a}}).
\newblock Distribution of the electron pitch-angle diffusion rates in the
  radiation belts.
\newblock In \emph{{EGU} {General} {Assembly} {Conference} {Abstracts}},
  volume~14, page 5192.

\bibitem[{Artemyev \emph{et~al.}(2012{\natexlab{b}})Artemyev, Agapitov,
  Breuillard, Krasnoselskikh, and Rolland}]{artemyev_electron_2012}
Artemyev, A., Agapitov, O., Breuillard, H., Krasnoselskikh, V., and Rolland, G.
  (2012{\natexlab{b}}).
\newblock Electron pitch-angle diffusion in radiation belts: {The} effects of
  whistler wave oblique propagation.
\newblock \emph{Geophysical Research Letters}, \textbf{39}(8).

\bibitem[{Bortnik \emph{et~al.}(2008)Bortnik, Thorne, and
  Meredith}]{bortnik_unexpected_2008}
Bortnik, J., Thorne, R.~M., and Meredith, N.~P. (2008).
\newblock The unexpected origin of plasmaspheric hiss from discrete chorus
  emissions.
\newblock \emph{Nature}, \textbf{452}(7183), 62--66.
\newblock ISSN 0028-0836.
\newblock
  \urlprefix\url{http://www.nature.com/nature/journal/v452/n7183/full/nature06741.html}.

\bibitem[{Breuillard \emph{et~al.}(2012)Breuillard, Mendzhul, and
  Agapitov}]{breuillard_effects_2012}
Breuillard, H., Mendzhul, D., and Agapitov, O. (2012).
\newblock Effects of equatorial chorus wave normal azimuthal distribution on
  wave propagation.
\newblock \emph{Advances in Astronomy and Space Physics}, \textbf{2}(2),
  167--172.

\bibitem[{Burton and Holzer(1974)}]{burton_origin_1974}
Burton, R.~K. and Holzer, R.~E. (1974).
\newblock The {Origin} and {Propagation} of {Chorus} in the {Outer}
  {Magnetosphere}.
\newblock \emph{{\textbackslash}jgr}, \textbf{79}, 1014--1023.

\bibitem[{Gallagher \emph{et~al.}(2000)Gallagher, Craven, and
  Comfort}]{gallagher_global_2000-1}
Gallagher, D.~L., Craven, P.~D., and Comfort, R.~H. (2000).
\newblock Global core plasma model.
\newblock \emph{Journal of Geophysical Research: Space Physics},
  \textbf{105}(A8), 18\,819--18\,833.
\newblock ISSN 2156-2202.
\newblock
  \urlprefix\url{http://onlinelibrary.wiley.com/doi/10.1029/1999JA000241/abstract}.

\bibitem[{Goldstein and Tsurutani(1984)}]{goldstein_wave_1984}
Goldstein, B.~E. and Tsurutani, B.~T. (1984).
\newblock Wave normal directions of chorus near the equatorial source region.
\newblock \emph{{\textbackslash}jgr}, \textbf{89}, 2789--2810.

\bibitem[{Gurnett \emph{et~al.}(1979)Gurnett, Anderson, Scarf, Fredricks, and
  Smith}]{gurnett_initial_1979}
Gurnett, D.~A., Anderson, R.~R., Scarf, F.~L., Fredricks, R.~W., and Smith,
  E.~J. (1979).
\newblock Initial results from the {ISEE}-1 and -2 plasma wave investigation.
\newblock \emph{Space Science Reviews}, \textbf{23}(1), 103--122.
\newblock ISSN 0038-6308, 1572-9672.
\newblock \urlprefix\url{http://link.springer.com/article/10.1007/BF00174114}.

\bibitem[{Hayakawa \emph{et~al.}(1984)Hayakawa, Yamanaka, Parrot, and
  Lefeuvre}]{hayakawa_wave_1984}
Hayakawa, M., Yamanaka, Y., Parrot, M., and Lefeuvre, F. (1984).
\newblock The wave normals of magnetospheric chorus emissions observed on board
  {GEOS} 2.
\newblock \emph{{\textbackslash}jgr}, \textbf{89}, 2811--2821.

\bibitem[{Helliwell(1967)}]{helliwell_theory_1967}
Helliwell, R.~A. (1967).
\newblock A theory of discrete {VLF} emissions from the magnetosphere.
\newblock \emph{Journal of Geophysical Research}, \textbf{72}(19), 4773--4790.
\newblock ISSN 2156-2202.
\newblock
  \urlprefix\url{http://onlinelibrary.wiley.com/doi/10.1029/JZ072i019p04773/abstract}.

\bibitem[{Mourenas \emph{et~al.}(2012{\natexlab{a}})Mourenas, Artemyev,
  Agapitov, and Krasnoselskikh}]{mourenas_acceleration_2012}
Mourenas, D., Artemyev, A., Agapitov, O., and Krasnoselskikh, V.
  (2012{\natexlab{a}}).
\newblock Acceleration of radiation belts electrons by oblique chorus waves.
\newblock \emph{Journal of Geophysical Research: Space Physics},
  \textbf{117}(A10), A10\,212.
\newblock ISSN 2156-2202.
\newblock
  \urlprefix\url{http://onlinelibrary.wiley.com/doi/10.1029/2012JA018041/abstract}.

\bibitem[{Mourenas \emph{et~al.}(2012{\natexlab{b}})Mourenas, Artemyev,
  Agapitov, and Krasnoselskikh}]{mourenas_consequences_2012}
Mourenas, D., Artemyev, A., Agapitov, O., and Krasnoselskikh, V.
  (2012{\natexlab{b}}).
\newblock Consequences of oblique chorus waves on the loss and acceleration of
  trapped electrons.
\newblock \emph{AGU Fall Meeting Abstracts}, \textbf{1}, 2374.

\bibitem[{Mourenas \emph{et~al.}(2012{\natexlab{c}})Mourenas, Artemyev, Ripoll,
  Agapitov, and Krasnoselskikh}]{mourenas_timescales_2012}
Mourenas, D., Artemyev, A.~V., Ripoll, J.-F., Agapitov, O.~V., and
  Krasnoselskikh, V.~V. (2012{\natexlab{c}}).
\newblock Timescales for electron quasi-linear diffusion by parallel and
  oblique lower-band chorus waves.
\newblock \emph{Journal of Geophysical Research: Space Physics},
  \textbf{117}(A6), A06\,234.
\newblock ISSN 2156-2202.
\newblock
  \urlprefix\url{http://onlinelibrary.wiley.com/doi/10.1029/2012JA017717/abstract}.

\bibitem[{Olson and Pfitzer(1974)}]{olson_quantitative_1974}
Olson, W.~P. and Pfitzer, K.~A. (1974).
\newblock A quantitative model of the magnetospheric magnetic field.
\newblock \emph{Journal of Geophysical Research}, \textbf{79}(25), 3739--3748.
\newblock ISSN 2156-2202.
\newblock
  \urlprefix\url{http://onlinelibrary.wiley.com/doi/10.1029/JA079i025p03739/abstract}.

\bibitem[{Omura \emph{et~al.}(1991)Omura, Matsumoto, Nunn, and
  Rycroft}]{omura_review_1991}
Omura, Y., Matsumoto, H., Nunn, D., and Rycroft, M.~J. (1991).
\newblock A review of observational, theoretical and numerical studies of {VLF}
  triggered emissions.
\newblock \emph{Journal of Atmospheric and Terrestrial Physics}, \textbf{53},
  351--368.

\bibitem[{Parrot \emph{et~al.}(2003)Parrot, Santolik, Cornilleau-Wehrlin,
  Maksimovic, Harvey, and {others}}]{parrot_source_2003}
Parrot, M., Santolik, O., Cornilleau-Wehrlin, N., Maksimovic, M., Harvey,
  C.~C., and {others} (2003).
\newblock Source location of chorus emissions observed by {Cluster}.
\newblock In \emph{Annales {Geophysicae}}, volume~21, pages 473--480.
\newblock
  \urlprefix\url{http://hal-obspm.ccsd.cnrs.fr/docs/00/32/92/34/PDF/angeo-21-473-2003.pdf}.

\bibitem[{SantolÃ­k \emph{et~al.}(2005)SantolÃ­k, Gurnett, Pickett, Parrot, and
  Cornilleau-Wehrlin}]{santolik_central_2005}
SantolÃ­k, O., Gurnett, D.~A., Pickett, J.~S., Parrot, M., and
  Cornilleau-Wehrlin, N. (2005).
\newblock Central position of the source region of storm-time chorus.
\newblock \emph{Planetary and Space Science}, \textbf{53}(1â€“3), 299--305.
\newblock ISSN 0032-0633.
\newblock
  \urlprefix\url{http://www.sciencedirect.com/science/article/pii/S0032063304001965}.

\bibitem[{Sazhin and Hayakawa(1992)}]{sazhin_magnetospheric_1992}
Sazhin, S.~S. and Hayakawa, M. (1992).
\newblock Magnetospheric chorus emissions: {A} review.
\newblock \emph{Planetary and Space Science}, \textbf{40}(5), 681--697.
\newblock ISSN 0032-0633.
\newblock
  \urlprefix\url{http://www.sciencedirect.com/science/article/pii/003206339290009D}.

\bibitem[{Stix(1962)}]{stix_theory_1962}
Stix, T.~H. (1962).
\newblock \emph{The {Theory} of {Plasma} {Waves}}.

\bibitem[{Trakhtengerts(1999)}]{trakhtengerts_generation_1999}
Trakhtengerts, V.~Y. (1999).
\newblock A generation mechanism for chorus emission.
\newblock \emph{Annales Geophysicae}, \textbf{17}, 95--100.

\bibitem[{Tsurutani and Smith(1974)}]{tsurutani_postmidnight_1974}
Tsurutani, B.~T. and Smith, E.~J. (1974).
\newblock Postmidnight chorus: {A} substorm phenomenon.
\newblock \emph{{\textbackslash}jgr}, \textbf{79}, 118--127.

\bibitem[{Tsurutani and Smith(1977)}]{tsurutani_two_1977}
Tsurutani, B.~T. and Smith, E.~J. (1977).
\newblock Two types of magnetospheric {ELF} chorus and their substorm
  dependences.
\newblock \emph{{\textbackslash}jgr}, \textbf{82}, 5112--5128.

\bibitem[{Yearby \emph{et~al.}(2011)Yearby, Balikhin, Khotyaintsev, Walker,
  Krasnoselskikh, Alleyne, and Agapitov}]{yearby_ducted_2011}
Yearby, K., Balikhin, M., Khotyaintsev, Y.~V., Walker, S., Krasnoselskikh, V.,
  Alleyne, H., and Agapitov, O. (2011).
\newblock Ducted propagation of chorus waves: {Cluster} observations.
\newblock \emph{Annales Geophysicae}, \textbf{29}(9), 1629--1634.

\end{thebibliography}
%\end{multicols}

\end{document}